\documentclass[aps,pre,twocolumn,superscriptaddress,nofootinbib]{revtex4-2}
\usepackage[section]{placeins}
\usepackage{ragged2e}
\usepackage[english]{babel}
\usepackage{amsmath}
\usepackage{amsfonts}
\usepackage{amssymb}
\usepackage{graphicx}
\usepackage{float}
\usepackage[utf8]{inputenc}
\usepackage{array}
\usepackage{physics}
\usepackage{xr-hyper}
\usepackage{hyperref}
\usepackage{xcolor}
\usepackage[normalem]{ulem}
\usepackage{booktabs}

\begin{document}
\title{Higher-Order OTOCs in Chaotic Quantum Maps: Ruelle–Pollicott Resonances and Asymptotic Freeness}

\author{Nicolás Gómez Brucks}
\affiliation{Departamento de F\'isica ``J. J. Giambiagi'' and IFIBA, FCEyN,
Universidad de Buenos Aires, 1428 Buenos Aires, Argentina}

\author{Ignacio Garc\'ia-Mata}
\affiliation{Instituto de Investigaciones F\'isicas de Mar del Plata (IFIMAR), Facultad de Ciencias Exactas y Naturales,
Universidad Nacional de Mar del Plata \& CONICET, 7600 Mar del Plata, Argentina}

\author{Diego A. Wisniacki}
\affiliation{Departamento de F\'isica ``J. J. Giambiagi'' and IFIBA, FCEyN,
Universidad de Buenos Aires, 1428 Buenos Aires, Argentina}

\begin{abstract}
Out-of-time-ordered correlators (OTOCs) measure the scrambling of
information and are a powerful diagnostic of chaos. In chaotic quantum
systems with a classical analog, their relaxation towards equilibrium is
dictated by the classical Ruelle--Pollicott resonances. We investigate
higher-order OTOCs in quantum maps, prototypical one-body systems with a
classical limit showing all the essential features of chaotic dynamics.
We demonstrate numerically that in the intermediate-time regime they
decay exponentially at a rate that depends only on the parity of the
order: odd-order OTOCs decay at the same rate as classical correlations,
given by the leading Ruelle--Pollicott resonance, and even-order ones at
twice that rate. Both are far slower than the naive expectation, under
which the rate would grow in proportion to the order. We then mimic the
effect of an environment by defining a coarse-grained superoperator whose
effect is to introduce decoherence into the system, and find that the expected
proportionality between rate and order is thereby restored. Finally, we
show that at long times each higher-order OTOC becomes dominated by a
single free cumulant, whose saturation value falls off inversely with the
dimension of the Hilbert space, signalling the emergence of asymptotic
freeness and highlighting the relevance of free probability theory as a
framework for the study of quantum chaos.
\end{abstract}

\maketitle
\date{\today}

\section{Introduction}
Classical chaos is defined through exponential sensitivity to initial conditions. However, since the very notion of a trajectory is meaningless in quantum mechanics and the Schrödinger equation is linear, a rigorous definition of chaos in quantum systems remains elusive. While early works in the field have relied on the study of spectral statistics, current research also focuses on finding dynamical probes of chaos. Out-of-time-ordered correlators (OTOCs) \cite{Maldacena2016, GarciaMata2023} have emerged as notable candidates for such a dynamical probe. They are a dynamical measure of scrambling of information in many-body quantum systems \cite{Swingle2018}, and their use as a chaos diagnostic has been demonstrated in many paradigmatic models \cite{Fortes2019}. 

In the last few years, there has been growing interest in studying higher-order generalizations of OTOCs. These involve a larger number of operators and multiple forward and backward time propagations, which make them more sensitive to the fine-grained features of information scrambling, such as the emergence of unitary designs \cite{RobertsYoshida2017}. 
In the context of Hamiltonian learning, where it was already demonstrated that the standard OTOC provides substantial advantages in quantum learning tasks compared to time-ordered correlators \cite{Schuster2023}, recent research \cite{Google2025} has shown through experimental implementations that higher-order OTOCs provide a more sensitive probe of the underlying quantum dynamics than standard OTOCs, making them more suited to Hamiltonian learning tasks. 
This interest in higher-order correlations extends to the study of thermalization.
Progress has been made to extend the original ETH ansatz, which describes two-point correlations, into a full ETH describing multi-time higher-order correlations \cite{FoiniKurchan2019}, which makes it possible to study higher-order hydrodynamics beyond the linear response regime. The validity of the full ETH ansatz has been demonstrated in paradigmatic models \cite{PappalardiFritzsch2025, FritzschProsen2025}.
All these developments have led to 
works deriving analytical results in quantum circuit models \cite{Fritzsch2026} and characterizing their decay time scales in prototypical quantum many-body systems \cite{Herrmann2026}.

The combinatorics of full ETH can be formalized using free probability \cite{ValliniPappalardi2026, PappalardiFoiniKurchan2022}, which extends traditional probability theory to noncommuting random variables such as random matrices. This simplifies the study of general multi-time correlation functions and makes it possible to calculate their long-time saturation values in terms of 
free cumulants  
which generalize the classical notion of cumulants.
More importantly, free probability introduces the concept of freeness, which extends the notion of statistical independence to noncommuting random variables \cite{Speicher1994}. It is expected that, in ergodic systems, time-evolved observables become asymptotically free from static observables in the thermodynamic or semiclassical limit. Hence, the decay of higher-order OTOCs at long times is interpreted, in the framework of free probability, as the emergence of freeness between time-evolved and static observables.
Furthermore, it offers a novel paradigm for studying chaotic dynamics, as it has been suggested that the emergence of freeness at late times can be viewed as a signature of chaos \cite{Camargo2025}.

In this work we study higher-order OTOCs defined for quantum maps on the torus, obtained by quantizing classical chaotic maps. These are the simplest systems exhibiting 
the fundamental features of chaos.
We show numerically that in the intermediate-time regime their decay is dictated by the classical Ruelle-Pollicott resonances (RPR), the isolated eigenvalues of the Perron-Frobenius operator that govern the decay of classical correlations\cite{Ruelle1986,Pollicott1985}. In particular, we show that the decay does not depend on the order of the OTOC but only its parity.
Even-order OTOCs decay at a rate given by the square of the leading RPR, like the standard OTOC, whereas odd-order OTOCs decay at a rate governed by its first power, just like the two-point correlator. 
These results partially agree with the analytical results of \cite{Fritzsch2026}, in which it was proved that all higher-order OTOCs decay as the standard OTOC in a minimal quantum circuit model.
In addition, we define a coarse-grained propagator to introduce decoherence into the systems and show that, in this case, the decay becomes order-dependent for all the maps we analyze.
Finally, we study the emergence of freeness at long times. 
We demonstrate asymptotic freeness in quantum maps in the fully chaotic regime by calculating numerically the long-time saturation values of free cumulants, showing that they vanish as the inverse of the Hilbert space dimension. These results was already proved numerically in Ref. \cite{ValliniPappalardi2026} for the kicked top.

The remainder of this paper is organized as follows. In Sec.~\ref{Out-of-time-ordered correlators} we review basic properties of the standard OTOC and define higher-order OTOCs. In Sec.~\ref{Quantum maps on the torus}, we introduce the quantum maps studied in this work. Sec.~\ref{Open quantum maps} describes the open quantum map formalism and the coarse-grained superoperator used to reveal the classical Ruelle-Pollicott resonances. In Sec.~\ref{Exponential decay in higher-order OTOCs}, we present our numerical results on the intermediate-time exponential decay of k-OTOCs under both unitary and coarse-grained evolution. Sec.~\ref{kotocs} summarizes the relevant aspects of free probability theory and its connection to higher-order correlators. In Sec.~\ref{freeness} we provide numerical evidence for the emergence of asymptotic freeness at long times. Finally, Sec.~\ref{conclusions} contains our conclusions.

\section{Out-of-time-ordered correlators}
\label{Out-of-time-ordered correlators}
The standard four-point OTOC is defined by the average of the product,
\begin{equation}
    \langle A(t)BA(t)B \rangle,
\end{equation}
where $A$ and $B$ are usually chosen as two traceless, Hermitian operators, and $A(t)$ is a time-evolved observable in the Heisenberg picture.
Its name stems from the unconventional time ordering of factors $A(t)$ and $B$, which gives the OTOC a richer structure compared to an ordinary two-point correlator, making it a sensitive probe of operator growth and information scrambling. 

Interest in OTOCs grew out of the study of a closely related quantity, the squared commutator (SC) $\langle [A(t),B]^{\dagger}[A(t),B] \rangle$. Originally studied in quasiclassical approaches to superconductivity \cite{Larkin1969}, the SC has attracted considerable attention because it exhibits an initial growth which is an indicator of operator spreading and that, for the so-called fast scramblers, is exponential $\sim e^{2\Lambda t}$ \cite{Maldacena2016} and defines a quantum Lyapunov exponent $\Lambda$. When a classical limit exists, this early-time exponential growth is governed by the Lyapunov exponent $\lambda_L$ of the classical system, a classical measure of chaos which governs the short-time exponential separation of trajectories in chaotic systems.
The agreement between the classical and the quantum Lyapunov exponent is verified in several works \cite{GarciaMata2018, Jalabert2018, Kurchan2016, Rozenbaum2017, ChenZhou2018, Rammensee2018, Morita2021}.
Consequently, early-time exponential growth of the SC has been proposed as a signature of chaos.
Nevertheless, the Lyapunov regime is not universal, as several examples are known of systems which are not chaotic but exhibit an initial exponential growth \cite{hashimoto2020exponential,chavez2019quantum,PilatowskiPRE2020,XuPRL2020,HummelPRL2019,kidd2021saddle} and, conversely, of chaotic systems that obey a power-law growth \cite{Fazio2018,Motrunich2018,Riddel2019,Fortes2019,craps2020lyapunov}.

More robust diagnostics of chaotic dynamics may be found in the long-time regime \cite{Fortes2019}, in which, after a time usually called the scrambling time $t^*$, the SC exhibits fluctuations around a plateau due to quantum interference effects. When a classical limit exists, the scrambling time coincides with the Ehrenfest time $t_E$, which is the time necessary for a wave packet to spread across the whole phase space \cite{Shepelyansky2020}. For a Hilbert space of dimension $d$ it is given by $t_E \approx \lambda_L^{-1} \ln (2\pi A d)$, with $A$ some classical action scale \cite{silvestrov2002ehrenfest}.

Alternatively, one may look for chaotic signatures in the OTOC, which is one of the terms appearing by expanding the commutator in the definition of the SC. 
In chaotic quantum systems with a weel definedclassical limit, it has been shown that OTOCs of traceless operators present an intermediate-time regime, after the Ehrenfest time $t_E$, where they decay exponentially at a rate given by the square of the leading RPR \cite{GarciaMata2018, Notenson2023}. The RPRs are the isolated eigenvalues of the Perron-Frobenius operator, and their moduli govern the decay of correlation functions in mixing systems.
After this exponential decay, the OTOCs reach a saturation regime where, due to finite-size effects, they fluctuate around a nonzero plateau, whose value decreases with the dimension of the Hilbert space \cite{Huang2019}.

For many-body systems without a classical limit, the definition of the quantum analog of the RPRs remains an active area of research. Different approaches have been proposed \cite{Prosen2002, Prosen2004, Prosen2007, Znidaric2024,duh2026quantum}, among which one line of research focuses on analyzing the Liouvillian spectrum of weakly open quantum systems \cite{Mori2024}: when a system is coupled to an environment, the relaxation towards equilibrium is governed by eigenvalues of the corresponding Lindblad superoperator. The slowest decay rate corresponding to these eigenvalues defines the so-called Liouvillian gap. Recent work has demonstrated that this Liouvillian gap dictates the decay of OTOCs in a paradigmatic many-body model \cite{Duarte2026}.
These results justify the large body of research focusing on OTOCs.
 
One may also consider higher-order generalizations of OTOCs in which multiple interleaved operators are evaluated at different times. 
Since from a statistical point of view, the ordinary OTOC is a four-point correlator, these may be viewed as higher moments encoding a richer statistical picture of the underlying operator dynamics.

In this work we study higher-order OTOCs defined by the expectation value over a maximally mixed state
\begin{equation}
    F_k(t) = \frac{1}{d}\operatorname{Tr}[(A(t)B)^k], \,\,\,\, k \in \mathbb{N},
    \label{k-OTOCs}
\end{equation}
which we refer to as $k$-OTOCs. Based on this definition, $F_1$ is a time-ordered correlator and $F_2$ is the standard four-point OTOC.

The focus of this work is to investigate and characterize the different dynamical regimes of $k$-OTOCs for quantum maps on the torus, focusing in particular on their intermediate-time regime and its relation to the classical RPRs. 

\section{Quantum maps on the torus}
\label{Quantum maps on the torus}
We consider quantum maps defined on the 2-torus. These are the quantized versions of the classical area-preserving maps, which are the simplest systems showing the essential characteristics of chaos. 

The 2-torus can be thought of as a periodically repeated square of unit length. This structure implies that its quantum counterpart has a discrete Hilbert space of dimension $d$ and an effective Planck constant given by $\hbar = 1/(2\pi d)$ \cite{Hannay1980}.

The evolution of the system then proceeds in discrete time steps $t \in \mathbb{N}$, governed by the $t$-th power of a $d \times d$ unitary operator $\mathcal{U}$.
All the maps we consider can be derived from kicked systems, and therefore their evolution operators can be written as the product of two noncommuting unitaries
\begin{equation}
    \mathcal{U} = T(\hat p)V(\hat q).
\end{equation}

We investigate three paradigmatic models of quantum chaos: the (Chirikov) standard map, the Harper map and the perturbed Arnold's cat map.
Each map depends on a parameter $K$. For the standard and Harper maps, increasing $K$ drives the dynamics from near-integrable, through mixed, to predominantly chaotic.

The standard map \cite{Chirikov1979, Chirikov2008} is defined through the equations
\[
\begin{aligned}
    p_{n+1} &= p_n + K_S \sin(2\pi q_n) \\
    q_{n+1} &= q_n + p_{n+1}
\end{aligned}
\qquad \text{mod } 1,
\]
which correspond to the Floquet operator
\begin{equation}
    \mathcal{U} = e^{-i\frac{\hat{p}^2}{2\hbar}}e^{-i\frac{K_S}{2\pi \hbar}\cos (2\pi \hat q)}.
\end{equation}
This map shows a transition from integrability to chaos determined by the chaos parameter $K_S$. For $K_S = 0$, the dynamics are integrable, with invariant tori completely foliating the phase space. At low values of the chaos parameter $K_S$, the Kolmogorov-Arnold-Moser (KAM) theorem guarantees that some invariant tori persist, resulting in the global stability of motion. At $K_S = K_c = 0.971635 \dots $ the last KAM curve breaks down and above $K_c$ unbounded diffusion in momentum takes place.
For $K_S \gtrsim 5$, the dynamics are essentially chaotic, although some islands may appear even at very large values of $K_S$.

The Harper map \cite{Artuso2011} is an approximation of the motion of a kicked charge in an external magnetic field. It is defined by the map
\[
\begin{aligned}
    p_{n+1} &= p_n - K_{H} \sin (2 \pi q_n) \\
    q_{n+1} &= q_n + K_{H} \sin(2 \pi p_{n+1})
\end{aligned}
\qquad \text{mod } 1.
\]
The corresponding Floquet operator is given by
\begin{equation}
    \mathcal{U} = e^{i\frac{K_{H}}{2\pi \hbar}\cos(2\pi \hat p)}e^{i\frac{K_{H}}{2\pi \hbar}\cos(2\pi \hat q)}.
\end{equation}
As for the standard map, the parameter $K_{H}$ determines the kind of dynamics. For $K_{H} < 0.11$, the system is integrable, while for $K_{H} > 0.63$ the dynamics are essentially chaotic, although small islands may appear at larger values of $K_H$.

Finally, we consider a perturbed version of Arnold's cat map, defined classically by
\[
\begin{aligned}
    p_{n+1} &= q_n + p_n - 2\pi K_A \sin(2 \pi q_n) \\
    q_{n+1} &= q_n + p_{n+1} + 2\pi K_A \sin(2 \pi p_{n+1})
\end{aligned}
\qquad \text{mod } 1.
\]
Upon quantization these result in the following Floquet operator
\begin{equation}
    \mathcal{U} = e^{-i\left[\frac{\hat{p}^2}{2 \hbar} - \frac{K_A}{\hbar}\cos(2 \pi \hat p)\right]} e^{i\left[\frac{\hat{q}^2}{2 \hbar} + \frac{K_A}{\hbar}\cos(2 \pi \hat q)\right]}.
\end{equation}
For $K_A \ll 1$, the Anosov theorem establishes that the map is structurally stable, uniformly hyperbolic, ergodic and mixing. The perturbation produces nontrivial RPRs, contrary to the unperturbed version of Arnold's cat map, which is an example of a maximally mixing map, meaning that all its RPRs are equal to zero.
Moreover, the RPR spectrum is very sensitive to the values of the perturbation, meaning that by sweeping through a short range of values of $K_A$ we can obtain a rich diversity of decay rates.

To reveal the classical features underlying the quantum dynamics, in the next section we introduce a coarse-grained propagator that 
simulates the effect of a diffusive environment coupled to the system.

\section{Open quantum maps}
\label{Open quantum maps}
The classical Perron-Frobenius (PF) operator is the time evolution operator for phase space densities. 
For an invertible map $M$ its action is defined by
\begin{equation}
    \mathcal{P}\rho(x) = \rho(M^{-1}x) = \int dy \, \delta(x-My)\rho(y).
\end{equation}
It is unitary on the space of square-integrable functions $L^2$, and therefore its spectrum lies on the unit circle in the complex plane.
When extended to a suitably chosen functional space allowing for distributions, the spectrum of the PF operator for uniformly hyperbolic, chaotic maps consists of an essential and a point spectrum $\{\lambda_i\}$, with $\lambda_0 = 1$ corresponding to the invariant density \cite{Blank2002}. 
These generalized, isolated eigenvalues are the RPRs. They are located strictly inside the unit disk, and, classically, their moduli determine the decay rate of correlation functions, thus quantifying the rate of mixing in a chaotic system.

The calculation of these resonances is generally very complicated, since we have to diagonalize an infinite-dimensional operator. Nonetheless, their values can be obtained by operating directly on the quantized maps, via a suitably constructed coarse-grained superoperator $\mathbf{S}_{\epsilon}$ \cite{GarciaMataSaraceno2004, GarciaMataSaraceno2005}, which depends on a parameter $\epsilon$ that sets the strength of the coarse-graining. 
This superoperator effectively mimics the effect of a diffusive environment that produces decoherence.
Its spectrum converges to that of a classical coarse-grained operator, whose spectrum in turn converges to the classical RPRs, as is proved in Ref. \cite{Nonnenmacher2003}, where it is shown that this correspondence emerges by first taking the semiclassical limit $d\rightarrow\infty$ at fixed coarse-graining parameter $\epsilon$ and subsequently taking the limit $\epsilon \longrightarrow 0$.
We refer to this quantum-classical relation as spectral correspondence.

We construct the superoperator $\mathbf{S}_{\epsilon}$ by a two-step process consisting of a unitary evolution $\mathbf{U}$ followed by the application of a noisy channel $\mathbf{D}_{\epsilon}$, resulting in the following prescription for the time evolution of observables in the Heisenberg picture
\begin{equation}
    A(t+1) = \mathbf{S}_{\epsilon} A(t) = \mathbf{D}_{\epsilon} \circ \mathbf{U} A(t) \equiv \sum_{\xi}c_{\epsilon}(\xi) T_{\xi}^{\dagger} \mathcal{U}^{\dagger} A(t) \mathcal{U} T_{\xi},
\end{equation}
where $\xi = (q, p)$ is a two-component phase-space vector, $T_{\xi}$ are the translation operators on the torus, which are unitary, and the coefficients $c_{\epsilon}(\xi)$ are positive and follow a quasi-Gaussian distribution. Their values are given by the two-dimensional discrete Fourier transform of,
\begin{equation}
    \tilde{c}(\mu, \nu) = e^{-\frac{\epsilon d}{2\pi}\left[\sin^2\left(\frac{\pi \mu}{d}\right) + \sin^2 \left(\frac{\pi \nu}{d}\right) \right]}.
\end{equation}
They define a convex combination of unitary operators and therefore a completely positive, trace-preserving and unital quantum channel. As such its largest-modulus eigenvalue is equal to one, which in general will not be simple, with the rest of its spectrum lying inside a circle of radius $r<1$.
If furthermore the ergodicity condition is met, then the map has one single eigenvalue equal to one.
Therefore, we may order its spectrum as follows,
\begin{equation}
    \lambda_0 = 1 > |\lambda_1| \geq |\lambda_2| \geq \dots
\end{equation}
These eigenvalues determine the long-time dynamics of the system.
For a suitable range of the coarse-graining parameter, the leading eigenvalues of $\mathbf{S}_{\epsilon}$ become approximately independent of both $d$ and $\epsilon$ and coincide with the classical Ruelle-Pollicott resonances, as shown in Ref. \cite{Nonnenmacher2003}.

\begin{figure}[htbp]
    \centering
    \includegraphics[width=1\linewidth]{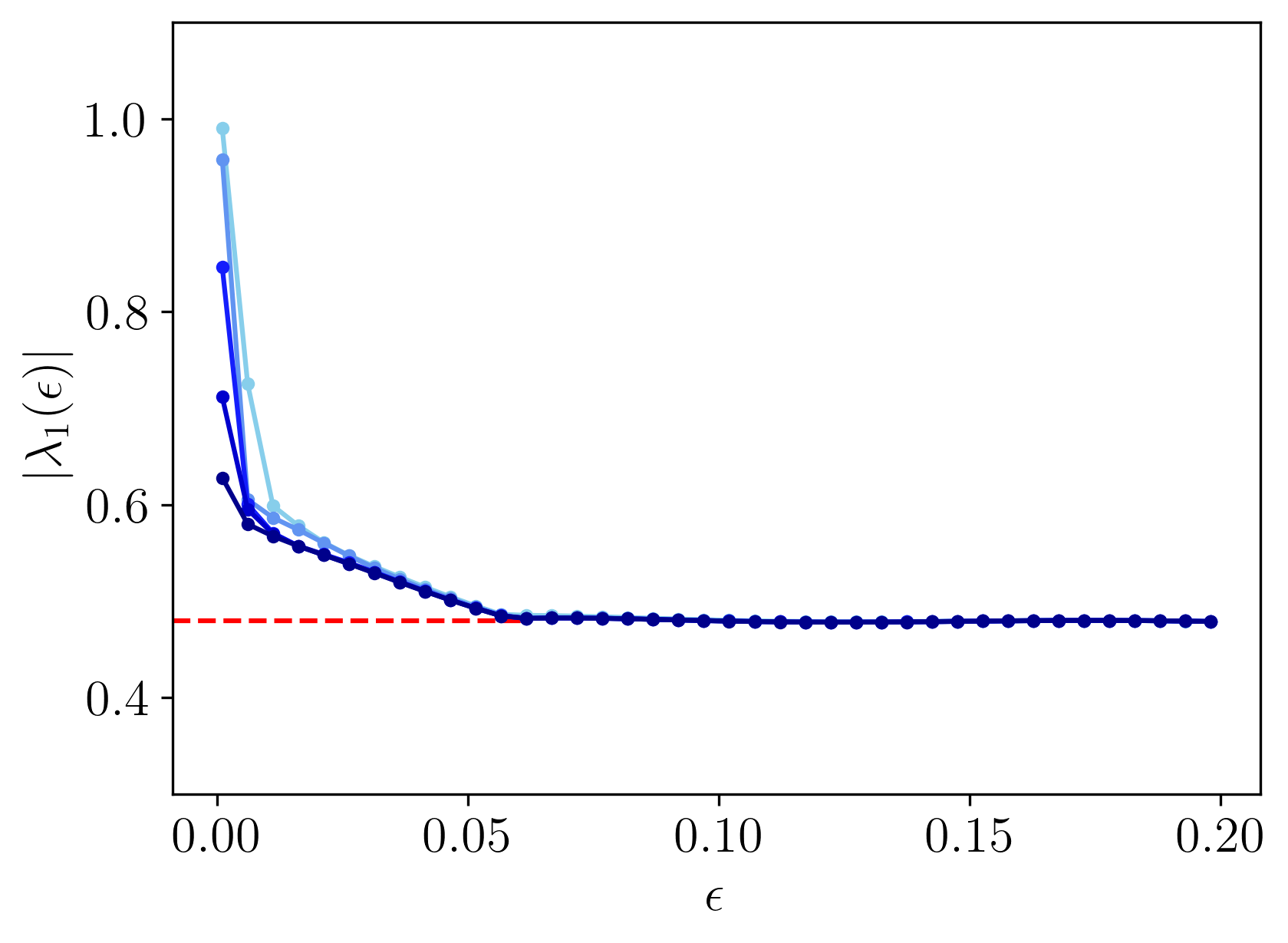}
    \caption{Modulus of the leading resonance $\lambda_1$ of the coarse-grained superoperator as a function of the coarse-graining parameter $\epsilon$ for the Harper map at $K_H = 0.87$. The curves correspond to Hilbert space dimensions  $d = 500, 1000, 2000, 3000$ and $d = 4000$, represented by progressively darker shades of blue. The dashed red line represents the mean value of $\lambda_1$ taken on a suitable interval where it stays constant, averaged over all Hilbert space dimensions.}
    \label{Leading resonance vs epsilon}
\end{figure}

Since we are only interested in the leading eigenvalues, we calculate the spectrum of $\mathbf{S}_{\epsilon}$ using the Arnoldi iteration, which provides a precise computation of the highest-modulus resonances. 
It is an iterative method that consists of building a Krylov subspace $K_m$ of dimension $m$ starting from an arbitrary initial vector that is, generically, not orthogonal to the leading eigenvector, and orthonormalizing it iteratively via a modified Gram-Schmidt procedure. 
This iteration produces a sequence of vectors which converges to the eigenvector corresponding to the eigenvalue of largest absolute value.
Furthermore, it yields an upper Hessenberg matrix $H_m$ whose eigenvalues, termed the Ritz values, converge, as $m$ grows, to the highest-modulus eigenvalues of $\mathbf{S}_{\epsilon}$ first, which is precisely the regime of interest here, since the leading Ruelle-Pollicott resonances are the eigenvalues of $\mathbf{S}_{\epsilon}$ of higher modulus. 
Figure~\ref{Leading resonance vs epsilon} shows the modulus of the leading nontrivial eigenvalue $\lambda_1$ of $\mathbf{S}_{\epsilon}$ for the Harper map at $K_H = 0.87$, plotted as a function of the coarse-graining parameter $\epsilon$ for several values of the Hilbert space dimension $d$. From the figure it is evident that $\lambda_1$ remains constant for a certain range of values of $\epsilon$ and $d$, providing numerical evidence for the spectral correspondence described before. 


The superoperator $\mathbf{S}_{\epsilon}$ is not normal, and therefore its left and right eigenoperators do not, in general, coincide.
\begin{figure*}[!t]
    \centering
    \includegraphics[width=1\linewidth]{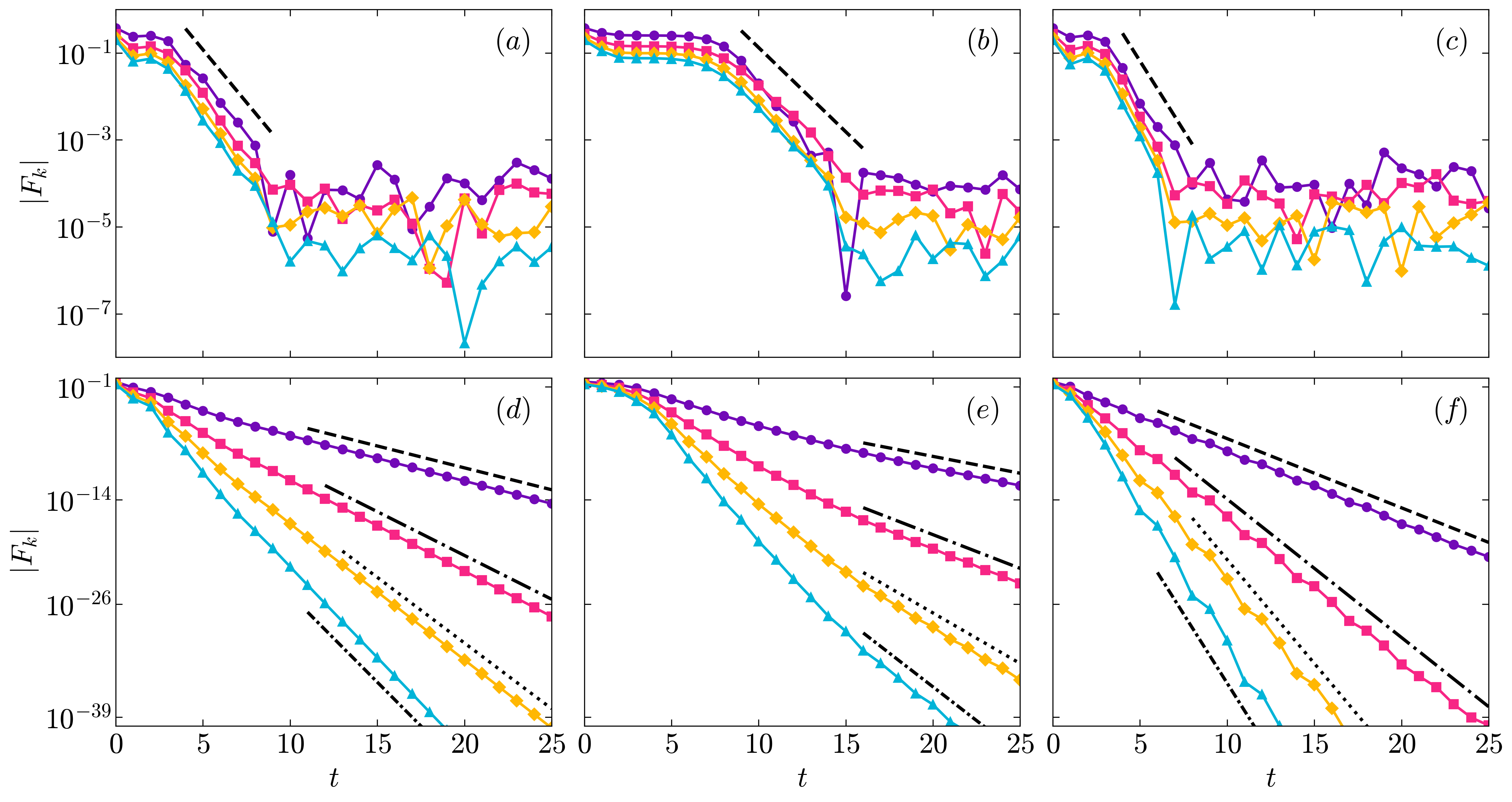}
    \caption{Exponential decay of even-order OTOCs under unitary (top row) and coarse-grained evolution (bottom row) for $\epsilon = 0.1$. From left to right, the columns show the standard map ($K_S=11.87$), the perturbed Arnold's cat map ($K_A=0.023$), and the Harper map ($K_H=0.87$). All panels display $F_k(t)$ for $P(t)P$ and $d=5000$: $F_2$ (purple circles), $F_4$ (pink squares), $F_6$ (yellow diamonds), and $F_8$ (cyan triangles). Dashed lines in panels (a)-(c) indicate the Ruelle-Pollicott decay $\propto |\lambda_1|^{2t}$; in panels (d)-(f), dashed, dash-dotted, dotted, and short dash-dotted lines indicate $\propto |\lambda_1|^{kt}$ for $k=2,4,6,$ and $8$, respectively. The leading resonance moduli are $|\lambda_1|\approx0.57$, $0.64$, and $0.48$ for the standard, perturbed cat, and Harper maps, respectively.}
    \label{even-order OTOCs}
\end{figure*}
They satisfy
\begin{equation}
    \mathbf{S}_{\epsilon}R_i = \lambda_i R_i \,\,\, \text{ and } \,\,\, \mathbf{S}^{\dagger}_{\epsilon}L_i = \lambda^{*}_i L_i,
\end{equation}
where the sets of right and left eigenoperators, $\{R_i\}$ and $\{L_i\}$, are biorthonormal with respect to the Frobenius inner product,
\begin{equation}
   d^{-1}\operatorname{Tr}(L_i^{\dagger}R_j) = \delta_{ij}.
\end{equation}

Assuming that these eigenoperators form a complete basis, the action of the coarse-grained superoperator on an operator $A$ admits the spectral decomposition
\begin{equation}
    \mathbf{S}_{\epsilon}A = \sum_i \lambda_i R_i d^{-1}\operatorname{Tr}(L_i^{\dagger}A),
\end{equation}
and its evolution after $t$ iterations is therefore given by
\begin{equation}
    A(t) = \mathbf{S}_{\epsilon}^t A = \sum_i \lambda_i^t R_i \operatorname{Tr}(L_i^{\dagger}A) = O(\lambda_1^t), 
    \label{spectral expansion A}
\end{equation}
for $t\gg 1$, where the last relation applies when the stationary contribution associated with $\lambda_0=1$ vanishes, since $\mathbf{S}_{\epsilon}$ is unital and trace-preserving, and therefore $L_0 \propto I$. 
This leads to the following asymptotic behavior for the $k$-OTOC,
\begin{equation}
    \langle(A(t)B)^k \rangle = \sum_{i_1,\dots,i_k}c_{i_1\dots i_k}(\lambda_{i_1}\dots\lambda_{i_k})^t = O(\lambda_1^{kt}), 
    \label{exact expansion OTOC}
\end{equation}
for $t\gg 1$, where the coefficients $c_{i_1\dotsi_k}$ are given by the product of projections onto the left eigenoperators by a factor which depends on right eigenoperators,
\begin{equation}
    c_{i_1\dots i_k} = \langle L_{i_1}^{\dagger}A\rangle\dots\langle L_{i_k}^{\dagger}A\rangle \langle R_{i_1}B\dots R_{i_k}B\rangle,
    \label{exact coefficients}
\end{equation}
where $\langle \cdot \rangle = d^{-1}\operatorname{Tr}(\cdot)$.

As we show in the next section, this expectation is verified for the coarse-grained dynamics. Under unitary evolution, by contrast, the decay remains governed by the RPRs, but its rate depends only on the parity of the order $k$.

\section{Exponential decay in higher-order OTOCs}
\label{Exponential decay in higher-order OTOCs}
In this section we present our results for the intermediate-time regime of $k$-OTOCs defined for the quantum maps. We analyze both unitary dynamics and coarse-grained dynamics, implemented by means of the superoperator $\mathbf{S}_{\epsilon}$ defined in the previous section. 
\begin{figure*}[!t]
    \centering
    \includegraphics[width=1\linewidth]{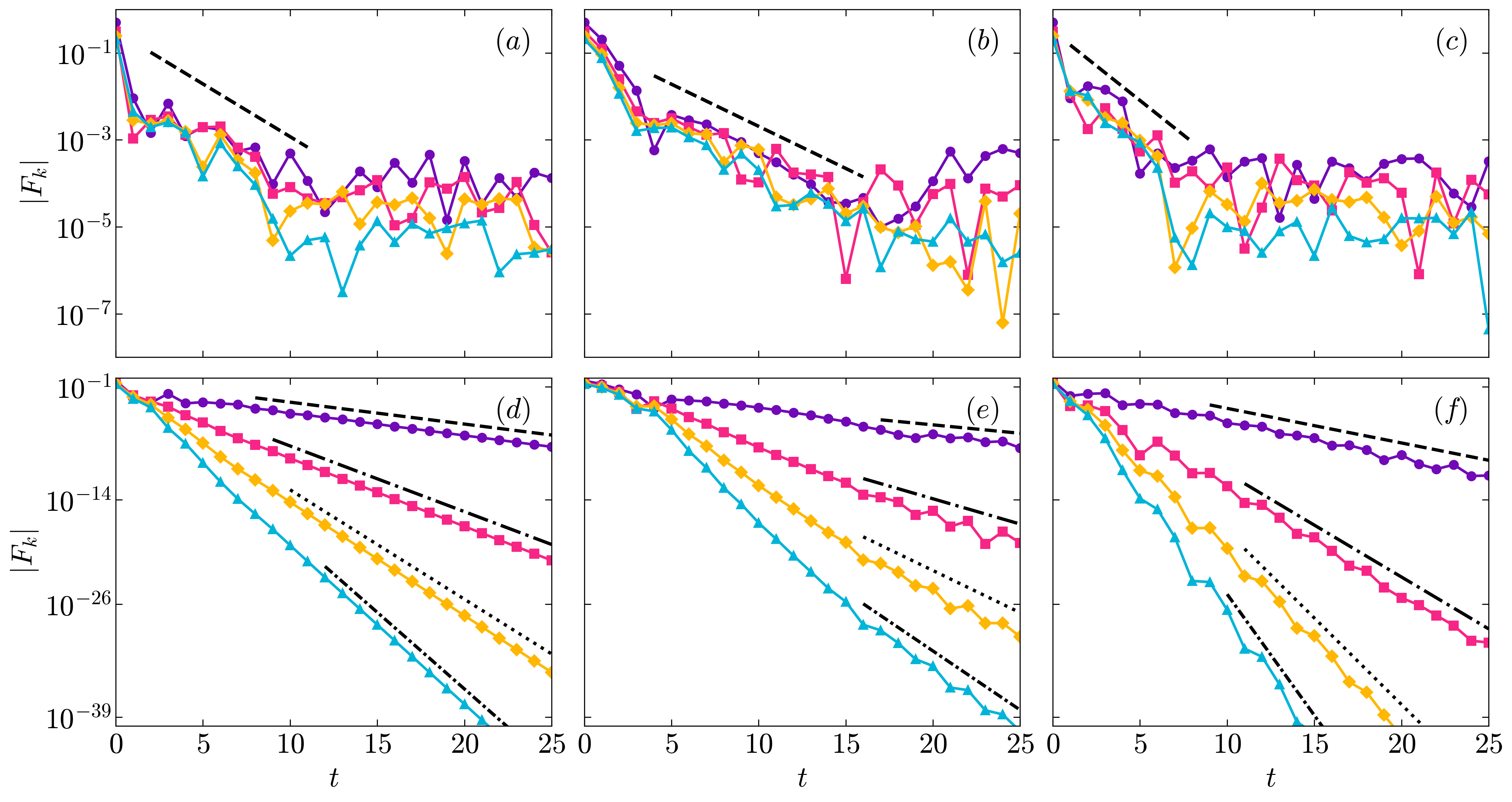}
    \caption{Exponential decay of odd-order correlators under unitary (top row) and coarse-grained evolution (bottom row) for $\epsilon = 0.1$. From left to right, the columns show the standard map ($K_S=11.87$), the perturbed Arnold's cat map ($K_A=0.023$), and the Harper map ($K_H=0.87$). All panels display $F_k(t)$ for $P(t)P$ and $d=5000$: $F_1$ (purple circles), $F_3$ (pink squares), $F_5$ (yellow diamonds), and $F_7$ (cyan triangles). Dashed lines in panels (a)-(c) indicate the Ruelle-Pollicott decay $\propto |\lambda_1|^t$; in panels (d)-(f), dashed, dash-dotted, dotted, and short dash-dotted lines indicate $\propto |\lambda_1|^{kt}$ for $k=1,3,5,$ and $7$, respectively. The leading resonance moduli are $|\lambda_1|\approx0.57$, $0.64$, and $0.48$ for the standard, perturbed cat, and Harper maps, respectively.}
    \label{odd-order OTOCs}
\end{figure*}
We work with $k$-OTOCs defined by Eq. \eqref{k-OTOCs}. The construction of suitable operators is not straightforward. As a consequence of the discreteness of the Hilbert space, the position and momentum operators are not well-defined on the 2-torus; therefore we need to construct operators which in the classical limit satisfy the correct commutation relations. 
An approximation can be constructed in terms of the Schwinger shift operators \cite{Schwinger1960} $U$ and $V$, which in the position basis are given by,
\begin{equation}
    U = \sum_{q \in \mathbb{Z}_d} e^{i\frac{2 \pi q}{d}} |q \rangle \langle q | \,\, \text{ and } \,\, V = \sum_{q \in \mathbb{Z}_d} |q + 1\rangle \langle q |.
\end{equation}

These operators are unitary, and their eigenvalues are the $d$-th roots of unity.
Furthermore, they satisfy the Weyl relation
\begin{equation}
    UV = e^{i\frac{2 \pi}{d}}VU.
\end{equation}
From them, we can construct the traceless, Hermitian operators $X$ and $P$,
\begin{equation}
    X = \frac{U - U^{\dagger}}{2i}, \,\,\, P = \frac{V - V^{\dagger}}{2i}.
\end{equation}
In the rest of this work we will consider different combinations of these operators to study the dynamics of $k$-OTOCs.

We show results for the standard map at $K_S = 11.87$, for the Harper map at $K_H = 0.87$ and for the perturbed Arnold's cat map at $K_A = 0.023$.
For the standard and Harper maps we have verified that no islands were visible, as it is known that regular regions may yield a slower, non-exponential decay of the OTOC \cite{Notenson2023}.

We find that, under unitary dynamics, after the Ehrenfest time, all even-order OTOCs decay at approximately the same rate, given by the square of the leading RPR $\propto |\lambda_1|^{2t}$. This can be seen in the first row of Fig.~\ref{even-order OTOCs}, where we show the intermediate-time regime of OTOCs $F_2, F_4, F_6$ and $F_8$ obtained by numerical simulations for all three maps for the operator $P(t)P$. 

As the unitary Floquet operator is replaced by the coarse-grained superoperator $\mathbf{S}_{\epsilon}$, we observe that all the decay rates become $k$-dependent, following $\propto |\lambda_1|^{kt}$. This is shown in the second row of Fig.~\ref{even-order OTOCs}.

For OTOCs of odd order, decoherence again restores the classical decay $\propto |\lambda_1|^{kt}$. Under unitary evolution, in contrast to the even-order case, we find a different intermediate-time behavior: the relaxation rate is dictated by $\propto |\lambda_1|^t$ for all orders. 
These results are shown in Fig.~\ref{odd-order OTOCs}, where we show the results of numerical evaluation of OTOCs $F_1, F_3, F_5$ and $F_7$ for all three maps for the operator $P(t)P$.

Notably, our results are partly in agreement with those of Ref.~\cite{Fritzsch2026}, where all higher-order OTOCs are found to decay at the same rate $\propto \lambda^{2t}$ in a minimal quantum circuit model. Nonetheless, in our case odd-order OTOCs decay at a slower rate dictated by the first power of the RPR.
This partial agreement suggests that the underlying operator combinatorics plays an important role in the dynamics of $k$-OTOCs, and that it may be possible to extend such combinatorial approaches to systems with a richer phase-space structure, such as the quantum maps studied here.

\begin{figure*}[!t]
    \centering
    \includegraphics[width=0.75\linewidth]{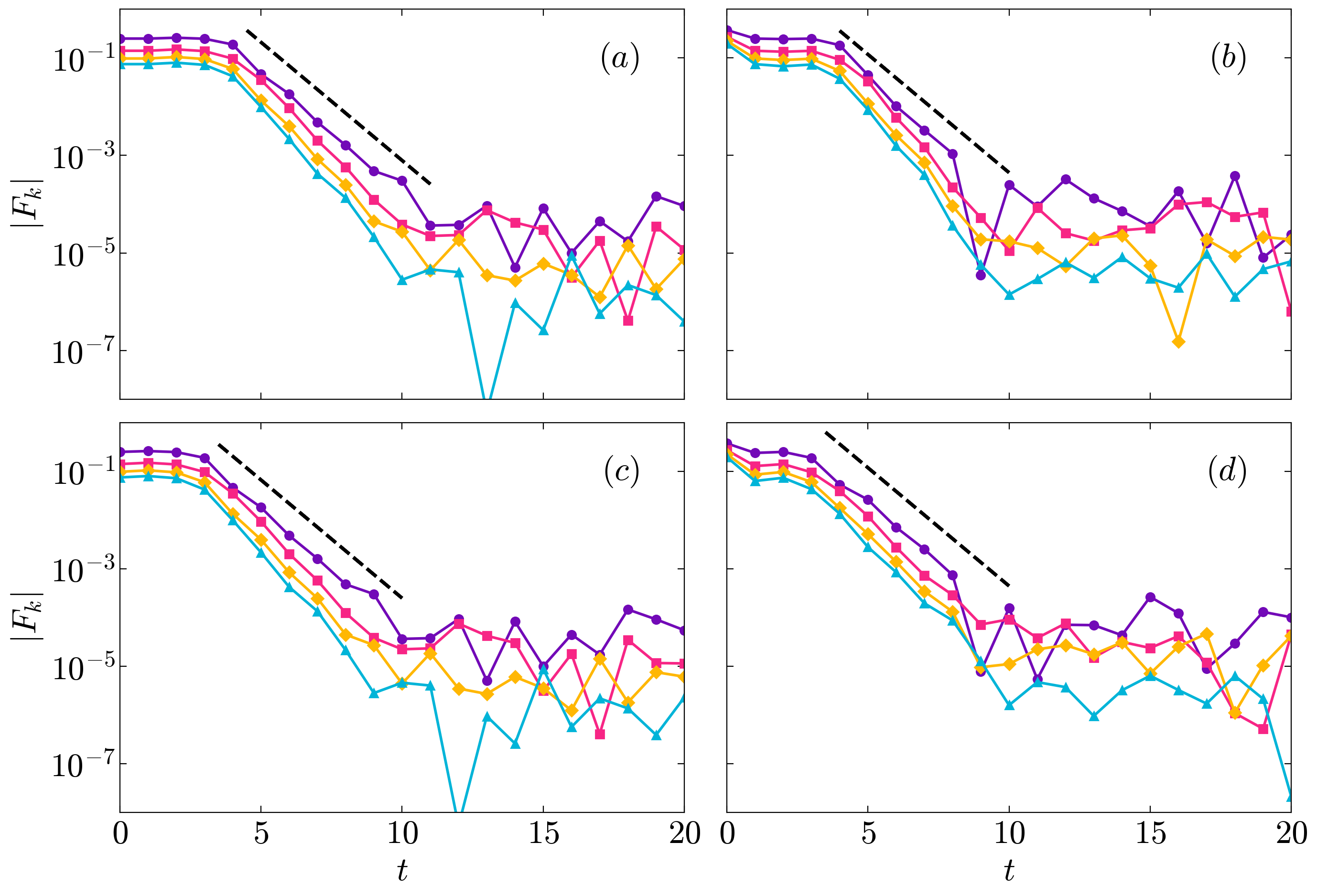}
    \caption{Comparison of even-order OTOCs for the four different combinations of operators $X$ and $P$ for the standard map at $K_S = 11.87$, $d = 5000$. Panels (a)–(d) show the results for $X(t)P$, $X(t)X$, $P(t)X$ and $P(t)P$, respectively. Each panel shows the OTOCs $F_2$ (purple circles), $F_4$ (pink squares), $F_6$ (yellow diamonds) and $F_8$ (cyan triangles). The dashed black lines indicate the common intermediate-time decay  $\propto |\lambda_1|^{2t}$, where $|\lambda_1| \approx 0.57$.}
    \label{Four operators comparison}
\end{figure*}

We obtained analogous results for all four possible combinations of operators $X$ and $P$, finding the same dichotomy in the even/odd-order behavior and the same relation to the Ruelle-Pollicott resonances. In fig. \ref{Four operators comparison} we compare the decays of OTOCs $F_2, F_4, F_6$ and $F_8$ for the four operators $X(t)P, X(t)X, P(t)X$ and $P(t)P$, for the standard map at $K_S = 11.87$, $d = 5000$. 
It can be seen that in all four cases the decay is governed by the leading RPR.
Analogous results were obtained for odd-order OTOCs for all four pairs of operators.

To conclude this section, we mention that in some cases it can be seen that the short time behavior of the coarse-grained dynamics may exhibit deviations from the asymptotic decay governed by the leading RPR. This is due to  a substantial overlap of $X$ and $P$ with subleading eigenoperators of  $\mathbf{S}_{\epsilon}$. We explain further this phenomenon in App. \ref{Appendix}.

\section{Higher-order OTOCs and free probability}
\label{kotocs}
As was pointed out in the introduction, the combinatorics of free probability provides the formalism to study multi-time correlation functions. 
In this section we summarize basic definitions of free probability, focusing on those topics more pertinent to our present work. A more general introduction to the topic may be found in Refs.~\cite{Speicher2009, MingoSpeicher2017}.

For two noncommutative random variables $A$ and $B$, the definition of free independence is given by \cite{Speicher2019},
\begin{equation}
    \begin{split}
        &\varphi\Bigl[(p_1(A)-\varphi(p_1(A)))(q_1(B)-\varphi(q_1(B)))\cdots
        \\
        &\,(p_n(A)-\varphi(p_n(A)))(q_n(B)-\varphi(q_n(B)))\Bigr] = 0.
    \end{split}
    \label{formal definition of free independence}
\end{equation}
This holds for all polynomials $p_i, q_i$ and all $n$. Here $\varphi(\cdot)$ denotes the expectation value and $p_i, q_i$ are polynomials in one variable.

The concept of free independence is greatly simplified by expressing it in terms of the so-called free cumulants. Free cumulants are connected correlation functions which generalize the concept of classical cumulants. They are defined through the combinatorics of noncrossing partitions, one of the fundamental objects in the theory of free probability. Given a partition $\pi$ of the set $\{1, 2, \dots, n \}$, we say that $\pi$ is noncrossing if, given two blocks $B_i, B_j \in \pi$, there are no elements $p_i, q_i \in B_i$ and $p_j, q_j \in B_j$ such that $p_i < p_j < q_i < q_j$.
We denote the set of all noncrossing partitions of the set $\{1, 2, \dots, n \}$ as $NC(n)$.

Given $n$ noncommutative random variables, free cumulants are defined recursively through the moment-cumulant relation
\begin{equation}
    \varphi(A_1\dots A_n) = \sum_{\pi \in NC(n)}\kappa_{\pi}(A_1, \dots  , A_n),
    \label{cumulants from moments}
\end{equation}
where 
\begin{equation}
    \kappa_{\pi}(A_1, \dots , A_n) = \prod_{b \in \pi}\kappa_{|b|}(A_{b(1)}, \dots ,A_{b(|b|)})
    \label{kappa pi}
\end{equation}
is the product, over blocks $b \in \pi$, of free cumulants $\kappa_{|b|}$.
The subscript $|b|$ indicates the size, i.e., the cardinality of each block $b$ of the partition $\pi$, and the notation $A_{b(i)}$ refers to the $i$-th element of the block $b$. 
Free cumulants appearing as factors in Eq. \eqref{kappa pi} are termed mixed free cumulants whenever at least one of the entries differs from the others.
Notably, an equivalent definition of free independence is provided by the vanishing of all mixed free cumulants.

For the case of two $d \times d$ random matrices $A, B$, they are said to be asymptotically free in the limit $d \longrightarrow \infty$, if all their mixed free cumulants vanish,
\begin{equation}
    \lim_{d \rightarrow \infty} \kappa_{n}(A,B, \dots ) = 0 \,\,\,\,\,\,\,\, n \geq 2 .
    \label{definition_of_asymptotic_freeness}
\end{equation}

Of particular relevance here is the fact that, given two deterministic (i.e., non-random) $d \times d$ matrices $A$ and $B$ and a Haar random unitary $U$, the matrices $U^{\dagger}AU$ and $B$ are asymptotically free \cite{MingoSpeicher2017}. Since we expect, for a chaotic system, the time evolution operator $\mathcal{U}(t)$ to mimic a Haar-random unitary, the time-evolved operator $A(t)$ should become asymptotically free from the static operator $B$.

Since we work with systems of finite size, exact asymptotic freeness cannot be expected. Instead, we test whether the mixed free cumulants vanish with increasing $d$ \eqref{definition_of_asymptotic_freeness}.

Eq. \eqref{cumulants from moments} simplifies when considering the operators $X$ and $P$ defined in section \ref{Exponential decay in higher-order OTOCs}, since they are traceless, and therefore $\kappa_1(X) = \kappa_1(P) = 0$.
Moreover, all their odd-order free cumulants vanish. This is due to the fact that all three maps are invariant under the parity transformation $q \rightarrow -q$ and $p \rightarrow -p$, which is implemented in their quantized counterparts by the operator $\Pi$ defined by
\begin{equation}
    \Pi |q\rangle = |-q\rangle,
\end{equation}
where $|q\rangle$ denotes the position basis.

The evolution operator is invariant under the action of $\Pi$,
\begin{equation}
    [\mathcal{U}, \Pi] = 0,
\end{equation}
while operators $X$ and $P$ transform as
\begin{equation}
     \Pi X \Pi = -X \,\, \text{ and } \,\, \Pi P \Pi = -P.
\end{equation}
Since odd-order free cumulants are a sum of products of moments, where every product contains at least one odd-length moment, this symmetry implies the vanishing of all odd-order free cumulants.
Thus, the following equations relate the moments and free cumulants of $X$ and $P$,
\begin{align}
    \kappa_{odd} &= 0 \nonumber \\
    \kappa_2 &= \langle X(t)P \rangle \nonumber \\
    \kappa_4 &= \langle (X(t)P)^2 \rangle - 2 \kappa_2^2 \nonumber \\
    \kappa_6 &= \langle (X(t)P)^3 \rangle - 5 \kappa_2^3 - 6 \kappa_2 \kappa_4 \\
    \kappa_8 &= \langle (X(t)P)^4 \rangle - 4 \kappa_4^2- 8 \kappa_2 \kappa_6  \nonumber \\
    & \,\,\,\,\,\,\,\,- 28 \kappa_2^2 \kappa_4 - 14 \kappa_2^4.
    \label{OTOCs and free cumulants}
\end{align}
Here we abbreviate
\begin{equation}
    \kappa_k = \kappa_k(X(t),P, \dots ).
\end{equation}
We have replaced the usual notation $\varphi(\cdot)$ by the normalized trace $\langle \cdot \rangle = d^{-1} \operatorname{Tr}( \cdot)$ which we have used in the rest of this work.  
The relations \eqref{OTOCs and free cumulants} hold for any of the four possible pairs of operators $X$ and $P$. This is due to the cyclic invariance of the trace and to the fact that only even-order free cumulants appear, each corresponding to a block of even length, which forces both arguments to appear an equal number of times. Therefore, no distinct mixed free cumulants, like $\kappa_2(X(t),X(t))$, appear.

The onset of asymptotic freeness was recently studied in Ref. \cite{ValliniPappalardi2026} for the quantum kicked top, which is a paradigmatic example of a chaotic quantum system with a classical limit. Based on analytical predictions of the full ETH ansatz, the authors proved numerically the following scaling for alternating free cumulants at long times in the fully chaotic regime,
\begin{equation}
    \kappa_k = O(d^{-1}), \,\,\, \text{for} \,\, t \gg 1.
    \label{free cumulants at long times}
\end{equation}

This scaling implies that the products of two or more free cumulants appearing in Eqs. \eqref{OTOCs and free cumulants} are negligible with respect to the individual free cumulant $\kappa_{2k}$ at long times and in the semiclassical limit, and therefore,
\begin{equation}
    \langle (A(t)B)^k \rangle \approx \kappa_{2k}, \,\text{ for} \,\, t \gg 1.
\end{equation}

In the next section, following Ref. \cite{ValliniPappalardi2026}, we analyze alternating free cumulants for quantum maps and test numerically whether the same $O(d^{-1})$ and $\kappa_{2k}$ dominance hold.

\section{Long-time asymptotic freeness}
\label{freeness}
We numerically investigate the emergence of freeness in chaotic quantum maps, following the methodology of Ref. \cite{ValliniPappalardi2026}. Here we show results for the standard and Harper maps, since they possess a chaos parameter which allows us to study both mixed and fully chaotic dynamics, but analogous results were obtained also for the perturbed Arnold's cat map for the fully chaotic regime.

\begin{figure}[htpb]
    \centering
    \includegraphics[width=\linewidth]{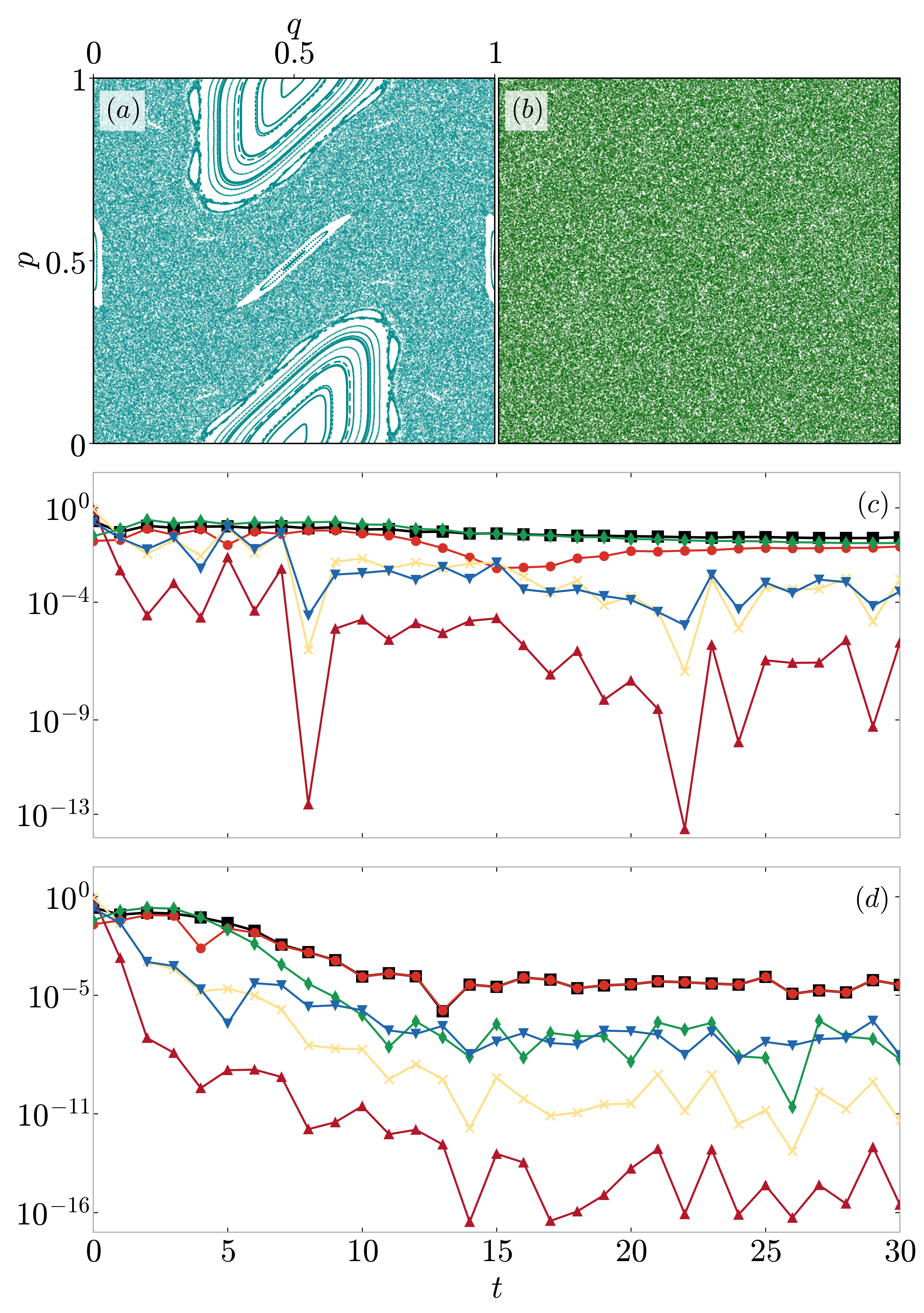}
    \caption{Phase portraits (top) and free-cumulant decomposition of the
    4-OTOC (bottom) for the standard map. The phase portraits and the corresponding free cumulant decomposition are shown for $K_S=2$ ((a) and (c), mixed phase space), and $K_S = 8$ ((b) and (d), fully chaotic). The black squares correspond to the 4-OTOC. The colored curves correspond to the terms of the decomposition: $\kappa_8$ (red circles), $4\kappa_4^2$ (green diamonds), $8\kappa_2\kappa_6$ (blue, downward triangles), $28\kappa_2^2\kappa_4$ (yellow crosses), and $14\kappa_2^4$ (crimson, upward triangles). The operator chosen is $P(t)P$, for Hilbert space dimension $d = 5000$.}
    \label{std free cumulant decomposition plots}
\end{figure}

Figure~\ref{std free cumulant decomposition plots} shows the free cumulant decomposition of the 4-OTOC for the standard map, together with the corresponding phase portraits, for two values of the chaos parameter $K_S$. These values correspond to two distinct dynamical regimes: $K_S= 2$ corresponds to mixed phase space and $K_S = 8$ to a fully chaotic regime.
In the fully chaotic case, the 4-OTOC is dominated by the leading free cumulant $\kappa_{8}$ already in the intermediate-time regime. 
In the mixed phase space case this behavior is not observed, indicating that asymptotic freeness is not reached when the system is not fully chaotic.
These results are consistent with the expectation that asymptotic freeness is achieved only when the dynamics are sufficiently chaotic, in agreement with Ref. \cite{ValliniPappalardi2026}.

\begin{figure}[htbp]
    \centering
    \includegraphics[width=\linewidth]{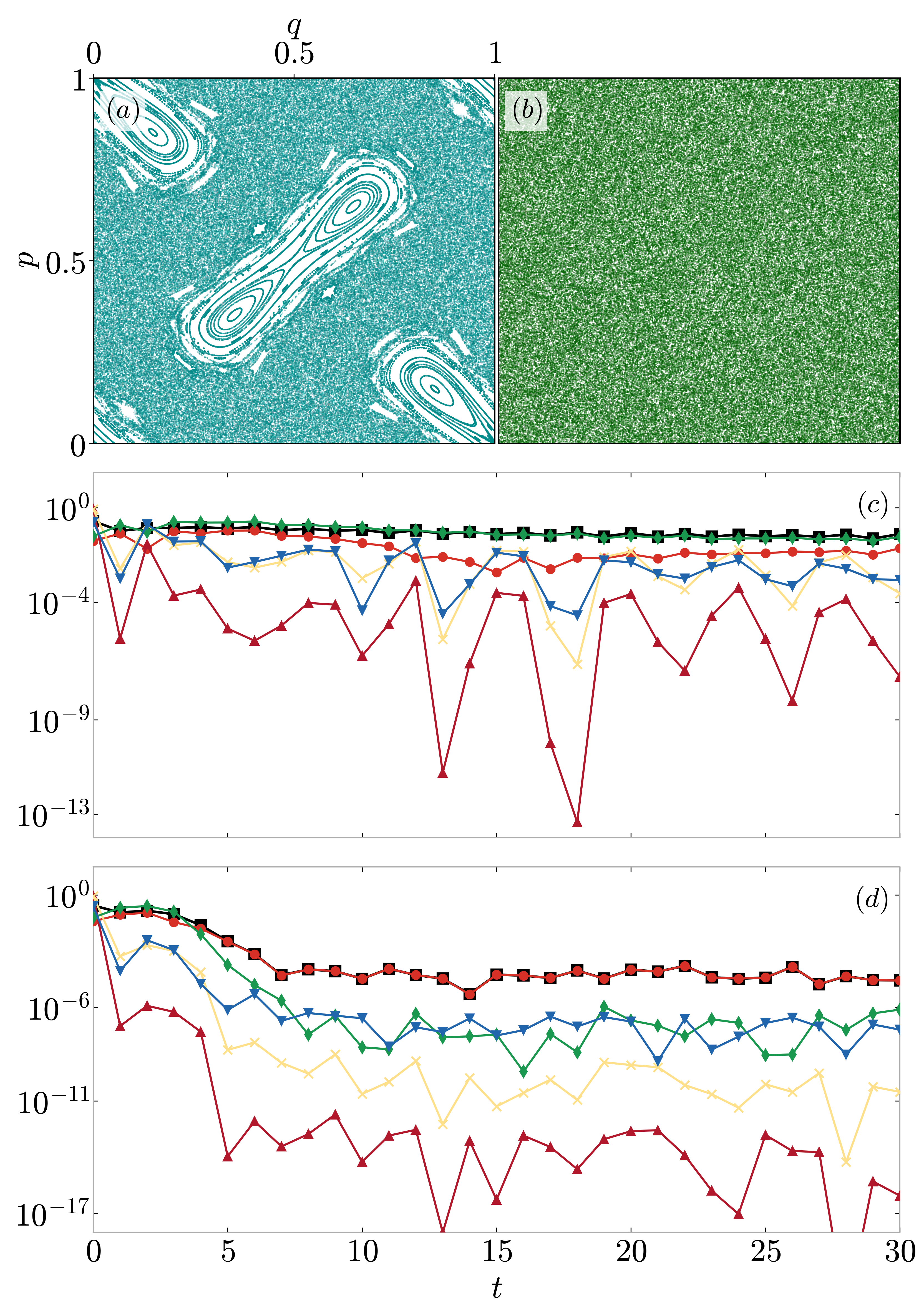}
    \caption{Phase portraits (top) and free-cumulant decomposition of the
    4-OTOC (bottom) for the Harper map. The phase portraits and the corresponding free cumulant decomposition are shown for $K_H=0.37$ (panels (a) and (c), mixed phase space) and $K_H = 0.87$ (panels (b) and (d), fully chaotic). The black squares correspond to the 4-OTOC. The colored curves correspond to the terms of the decomposition: $\kappa_8$ (red circles), $4\kappa_4^2$ (green diamonds), $8\kappa_2\kappa_6$ (blue, downward triangles), $28\kappa_2^2\kappa_4$ (yellow crosses), and $14\kappa_2^4$ (crimson, upward triangles). The operator chosen is $P(t)P$, for Hilbert space dimension $d = 5000$.}
    \label{har free cumulant decomposition plots}
\end{figure}

An  analogous behavior is observed for the Harper map in Fig. \ref{har free cumulant decomposition plots}. As $K_H$ is increased from $0.37$ to $0.87$, the classical phase space evolves from a mixed regime to a chaotic one. 
Correspondingly, the contribution of $\kappa_8$ becomes progressively dominant in the free-cumulant decomposition of the 4-OTOC. For $K_H = 0.87$, the agreement $F_4 \approx \kappa_8$ is already established at intermediate times, signaling the emergence of asymptotic freeness. At the lower value of $K_H = 0.37$, the remaining regular structures delay or prevent this dominance, consistently with the behavior found for the standard map. 

\begin{figure}
    \centering
    \includegraphics[width=1\linewidth]{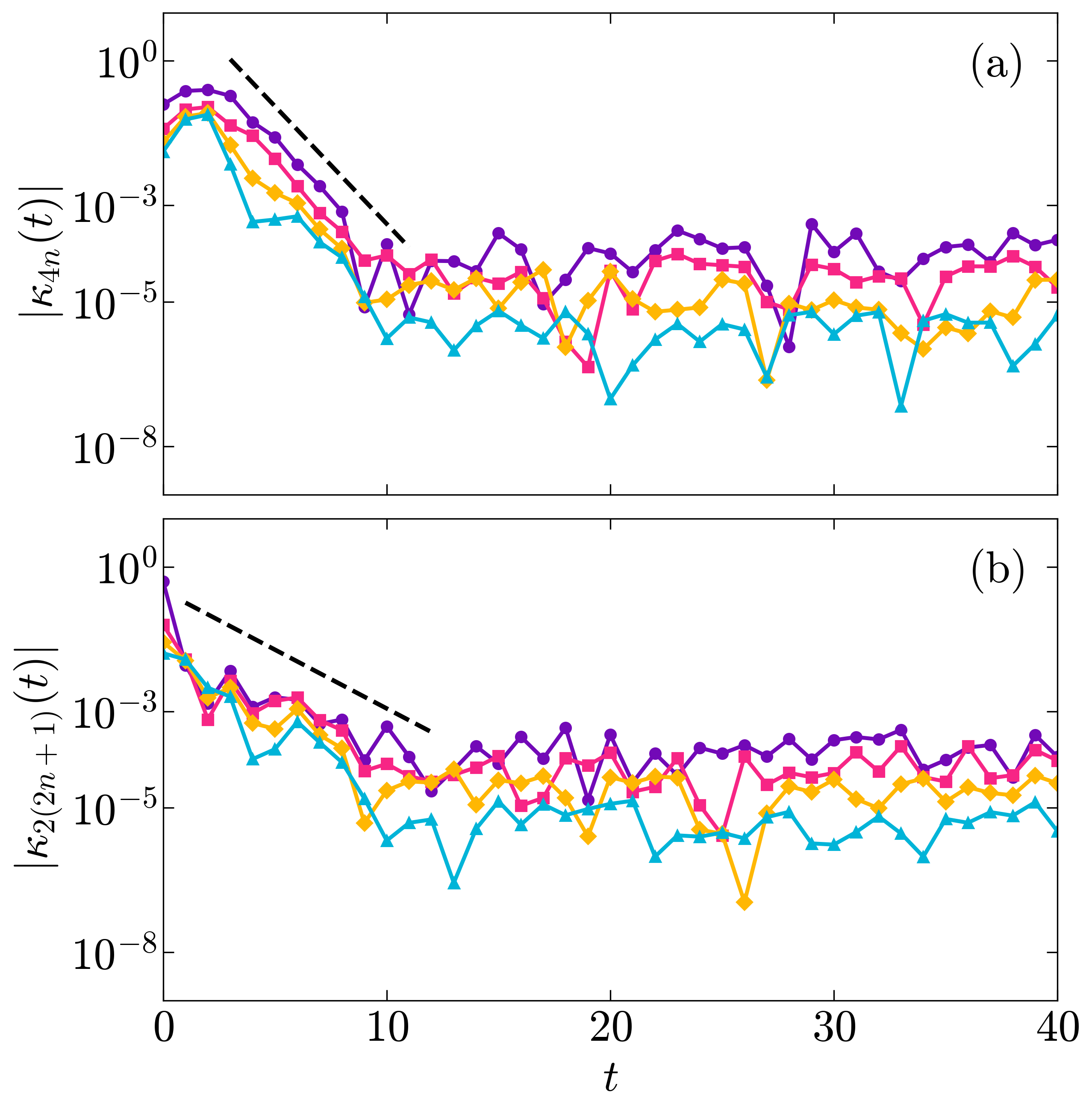}
    \caption{Even-order free cumulants for the standard map at $K_S = 11.87$ and $d = 5000$, for the operator $P(t)P$. Panel (a) shows the decay of free cumulants $\kappa_4$ (purple circles), $\kappa_8$ (pink squares), $\kappa_{12}$ (yellow diamonds) and $\kappa_{16}$ (cyan triangles), together with the decay predicted by the leading RPR $|\lambda_1|^{2t}$ where $|\lambda_1| \approx 0.57$, represented as a dashed black line. Panel (b) shows the free cumulants $\kappa_2$ (purple circles), $\kappa_6$ (pink squares), $\kappa_{10}$ (yellow diamonds) and $\kappa_{14}$ (cyan triangles). The black dashed line represents the Ruelle decay $|\lambda_1|^{t}$.}
    \label{free cumulants decay}
\end{figure}

The fact that the $k$-OTOC is dominated by the leading cumulant $\kappa_{2k}$ already at intermediate times indicates that, in the fully chaotic regime, the relaxation towards equilibrium of free cumulants is also governed by the leading RPR. This result is confirmed by numerical computation of free cumulants and is shown in Fig.~\ref{free cumulants decay}, where we plot the even-order free cumulants from $\kappa_2$ up to $\kappa_{16}$ for the standard map at $K_S = 11.87$ and $d = 5000$, together with the corresponding decays. It can be seen that the decay rate of free cumulants $\kappa_{2k}$ depends on the parity of $k$: for even $k$ they decay as $|\lambda_1|^{2t}$, while for odd $k$ as $|\lambda_1|^t$.

\begin{figure}[htbp]
    \centering
    \includegraphics[width=1\linewidth]{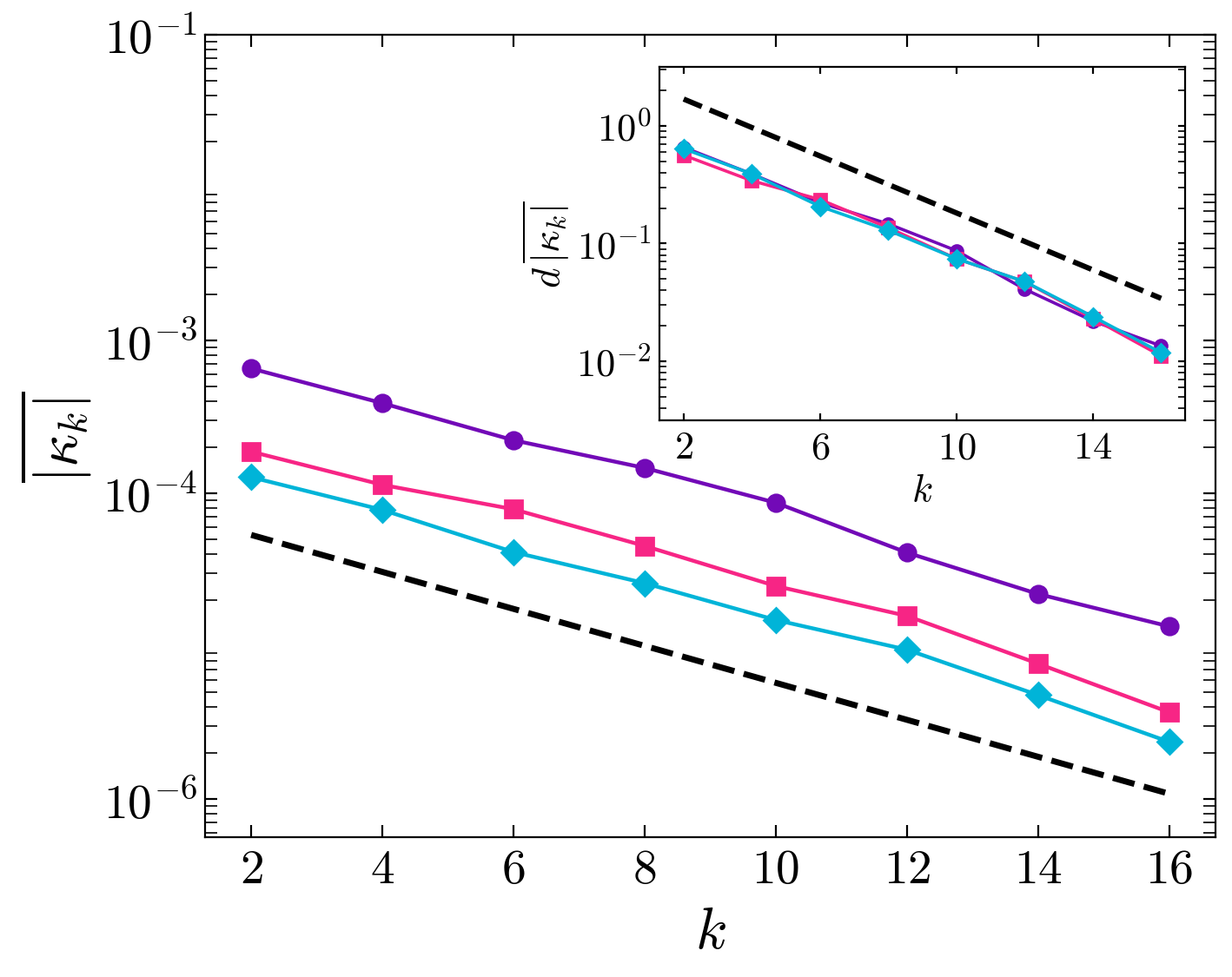}
    \caption{Mean saturation values of free cumulants for the standard map, $K_S = 8$, for three values of $d$: $d = 1000$ (purple circles), $d = 3000$ (pink squares) and $d = 5000$ (cyan diamonds). The operator chosen is $X(t)P$. The inset shows the mean saturation values multiplied by the Hilbert space dimension $d$. In both the main image and the inset the dashed black line corresponds to the exponential fit $\propto (0.76)^k$.}
    \label{Free cumulants plateaus}
\end{figure}

Finally, we test numerically the validity of the scaling \eqref{free cumulants at long times}. In Fig. \ref{Free cumulants plateaus} we plot the mean saturation values of the even-order free cumulants $\overline{|\kappa_{k}|}$ for completely chaotic dynamics as a function of the order $k$ and for different values of $d$, obtained by taking the average in the interval $t\in [30, 100]$. We have chosen to present the results for the standard map at $K_S = 8$. As is evident from the plot, the plateau value decays exponentially with the order $k$, and overall, its long-time dependence is given by,
\begin{equation}
    \overline{|\kappa_{k}|} \sim \frac{\left(0.76\right)^{k}}{d}, \text{ for even }  k.
\end{equation}
While the exact value determining the dependence on $k$ is operator-dependent, the dependence on the system size $d$ is the one predicted by equation \eqref{free cumulants at long times}.

We obtained equivalent results for the long-time behavior of free cumulants for the Harper and Arnold's cat map, arriving at the same scaling $O(d^{-1})$.
These results apply only when the dynamics are fully chaotic, while for mixed dynamics no analogous behavior was observed.

\section{Conclusions}
\label{conclusions}
\noindent
We have studied the decay of higher-order OTOCs in quantum maps on the
torus, in
which the classical Ruelle--Pollicott spectrum can be computed
independently and compared directly against the quantum dynamics.

Our main result is that the intermediate-time decay of $k$-OTOCs under
unitary evolution is governed by the leading Ruelle--Pollicott resonance
of the corresponding classical map, at a rate that does not grow with the
order but depends only on its parity: odd-order OTOCs decay as
$|\lambda_1|^{t}$, the same rate as ordinary two-point correlations,
while even-order ones decay as $|\lambda_1|^{2t}$. We have verified this
for three paradigmatic maps, and for the four combinations of the position and
momentum operators $X$ and $P$.

The even-order behavior is consistent with Ref.~\cite{Fritzsch2026},
where all OTOCs of order $k \geq 2$ were shown to decay as $\lambda^{2t}$
in an exactly solvable circuit model; the odd orders, which relax more
slowly, have no counterpart there. We stress that all the observables
used here are built from $X$ and $P$, which are odd under the parity
transformation that leaves the three maps invariant. 

By introducing decoherence, by coarse-graining the propagator of density matrices,
the classical
features of the dynamics are exposed, and the order-dependent decay 
$|\lambda_1|^{kt}$, for every OTOC considered, is restored. Deviations from the 
approach to the 
asymptotic rate is not immediate can be explained when the operators have a substantial
overlap with subleading eigenoperators of the coarse-grained
superoperator.

On the free-probability side, we found that the $k$-OTOC is dominated by
the single free cumulant $\kappa_{2k}$ already in the intermediate-time
regime, so that the free cumulants inherit the same parity-dependent
decay, and that their long-time saturation values scale as $O(d^{-1})$.
This confirms for quantum maps the behavior established in
Ref.~\cite{ValliniPappalardi2026} for the kicked top. Both features
appear only when the classical phase space is fully chaotic: for mixed
dynamics the surviving regular structures delay or prevent them, so the
onset of freeness tracks the chaoticity of the underlying classical map.

Taken together, these results tie the relaxation of higher-order quantum
correlations directly to the classical Ruelle--Pollicott spectrum, and
show that the approach to asymptotic freeness in systems with a classical
limit is exponentially fast at a rate set by that same spectrum. A
natural continuation is to ask how much of this survives in many-body
systems without a classical analog, where the role of the resonances is
played by the Liouvillian gap \cite{Mori2024, Duarte2026}, and whether
the parity dichotomy reported here can be resolved on the circuit-based
platforms where higher-order OTOCs have recently been
measured \cite{Google2025}.
\acknowledgments
We acknowledge support from Argentinian Agencia I$+$D$+$i (Grants No. PICT-2020-SERIEA-00740 and PICT-2020-SERIEA-01082). I.G-M received support from the French-Argentinian International Research Project \textit{Complex Quantum Systems} (COQSYS), funded by CNRS.
D.A.W.  received support from CONICET (Grant No. PIP 11220200100568CO), UBACyT (Grant No. 20020220300049BA) and PREI-UNAM.

\appendix
\section{Short-time deviaton from $\lambda_1$ decay}
\label{Appendix}
\begin{figure}[H]
    \centering
    \includegraphics[width=1.05\linewidth]{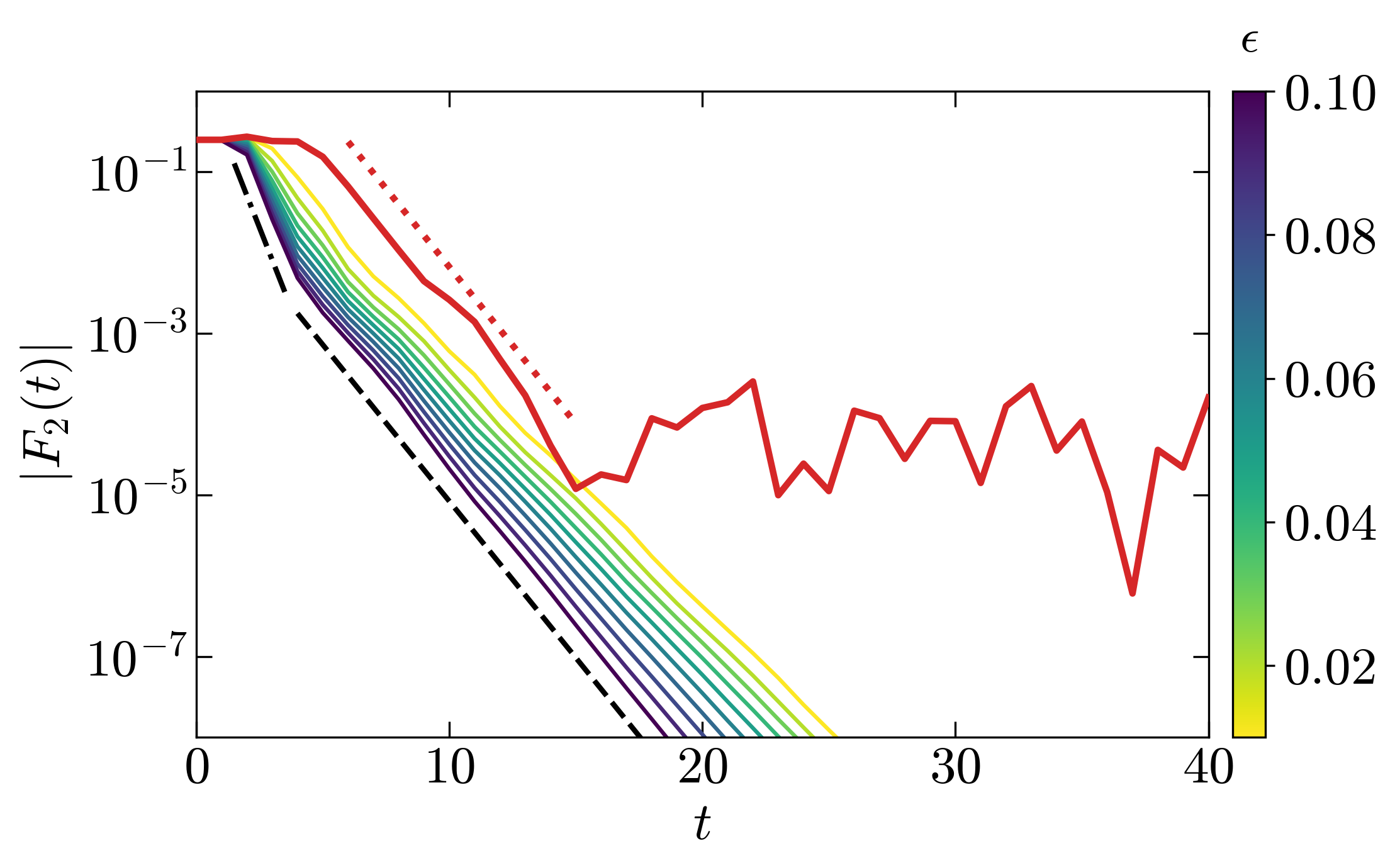}
    \caption{Decay of the 2-OTOC for the operator $X(t)P$ for the standard map at $K_S = 8$ and $d = 5000$. The red curve shows unitary evolution, with an intermediate-time decay $\propto |\lambda_1|^{2t}$, where $|\lambda_1| \approx 0.64$ is the leading Ruelle-Pollicott resonance. Coarse-grained evolution is color-coded along a smooth color gradient which indicates the value of the coarse-graining parameter $\epsilon$, swept through a range between $\epsilon = 0.01$ and $\epsilon = 0.1$. For larger values of $\epsilon$ we can observe two decay regimes, the first corresponding to smaller resonances which happen to have a significant overlap with the operator $X$.  The dashed black and dash-dotted lines indicate the reference decays  $\propto |\lambda_1|^{2t}$ and $\propto 0.40^{2t}$, respectively.}
    \label{Multiple Ruelle decay}
\end{figure}

As noted at the end of section \ref{Exponential decay in higher-order OTOCs}, the short-time behavior of the coarse-grained dynamics may exhibit deviations from the asymptotic decay governed by the leading RPR, as both even- and odd-order OTOCs may initially decay at a different rate. To investigate this phenomenon, we consider the 2-OTOC of the standard map at $K_S = 8$ for the operator $X(t)P$, shown in Fig.~\ref{Multiple Ruelle decay}, where this behavior is particularly manifest. 
There we compare the unitary and coarse-grained evolution for different values of the coarse-graining parameter $\epsilon$, and it can be seen that, for larger values of $\epsilon$, the initial decay is faster than that associated with the leading resonance $\lambda_1$, as the OTOC initially decays at a rate given by $\approx 0.40^{2t}$, before eventually decaying as the leading Ruelle resonance $|\lambda_1|^{2t} \approx 0.64^{2t}$.

\begin{figure}[htpb]
    \centering
    \includegraphics[width=1\linewidth]{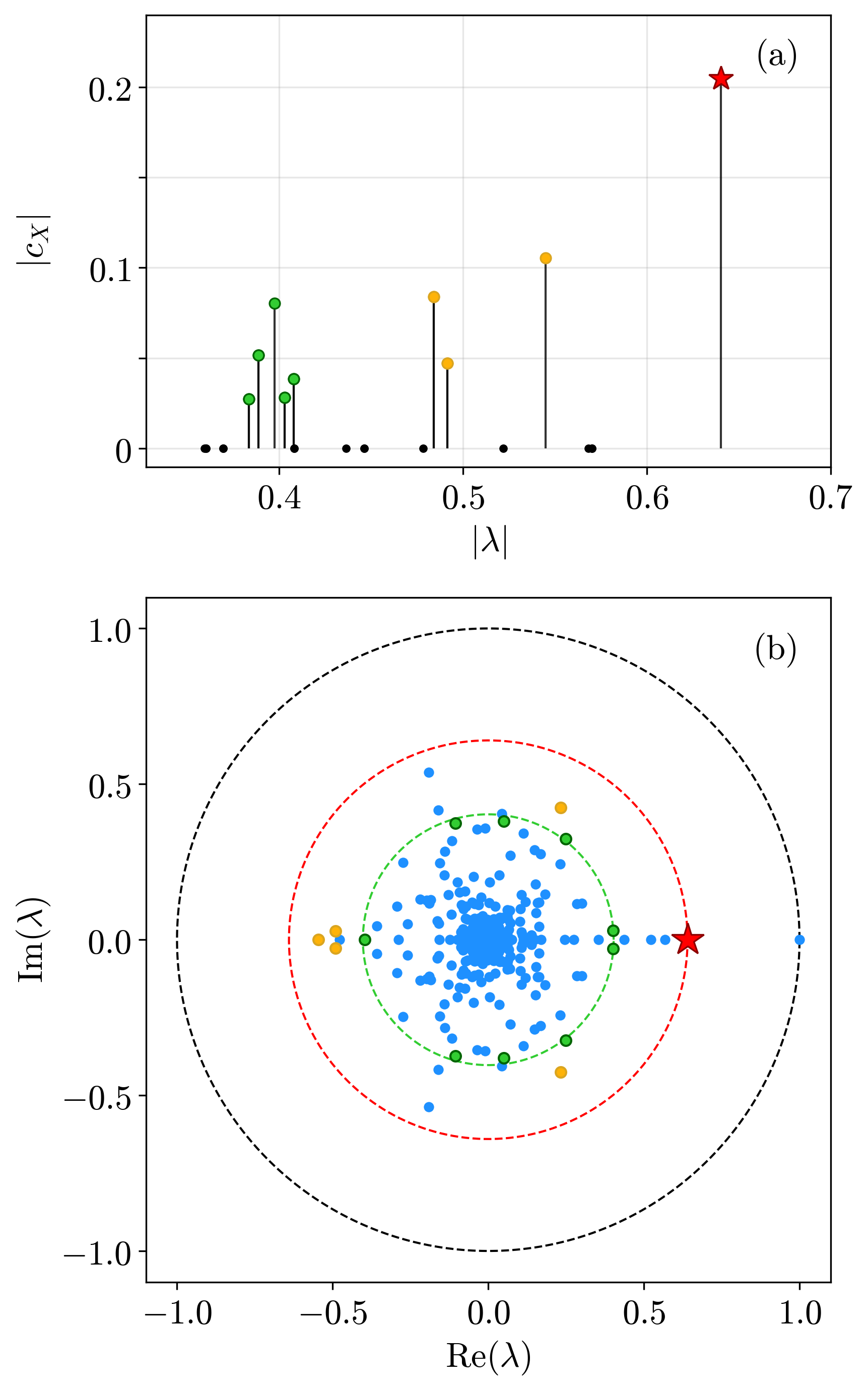}
    \caption{(a) Projection of the operator $X$ onto the first thirty  nontrivial left eigenoperators $L_i$ of the coarse-grained superoperator for the standard map at $K_S = 8$, $d = 1000$ and $\epsilon = 0.1$. The horizontal axis denotes the modulus of each eigenvalue. The vertical axis shows the projection coefficients $c_X = d^{-1} \operatorname{Tr} (L_i^{\dagger}X)$, which quantify the overlap of $X$ with each eigenmode. (b) Spectrum of the coarse-grained superoperator plotted in the complex plane. The dashed black circle denotes the unit circle. In both panels, the red stars denote the leading eigenvalue $\lambda_1$, the yellow circles the intermediate resonances $\lambda_5, \lambda_7$ - $\lambda_{10}$ and the green circles the cluster of resonances $\lambda_{17}$ up to $\lambda_{25}$. In panel (b) the dashed red circle identifies the leading resonance $\lambda_1$ while the dashed green circle passes through $\lambda_{20}$, where $|\lambda_{20}| \approx 0.40$.}
    \label{Overlaps and spectrum}
\end{figure}
\noindent

This initial behavior is caused by the operators $X$ and $P$ having a substantial overlap with subleading eigenoperators of $\mathbf{S}_{\epsilon}$, causing eigenvalues smaller in modulus to dominate the initial dynamics, before the leading, slowest resonance starts to dominate at larger times.
To prove this, we calculate, via the Arnoldi iteration, the spectrum of the coarse-grained operator $\mathbf{S}_{\epsilon}$ and the projections $c_X = d^{-1} \operatorname{Tr} (L_i^{\dagger}X)$ of resonances up to $\lambda_{30}$, in order to calculate the exact expansion \eqref{exact expansion OTOC} up to $\lambda_{30}$.
We made this analysis for the 2-OTOC only, and not for higher-order OTOCs, since the number of coefficients \eqref{exact coefficients} grows exponentially with the order of the OTOC, making a calculation of the exact expansion \eqref{exact expansion OTOC}, which involves multiple matrix products, computationally expensive.
As the coefficients \eqref{exact coefficients} appearing in the spectral expansion of the OTOC are the product of projections $c_X$ and a factor depending only on the right eigenoperators, only those eigenvalues with a nonzero overlap $c_X$ will contribute to the OTOC expansion \eqref{exact expansion OTOC}.
Panel (a) of Fig.~\ref{Overlaps and spectrum} shows the projections $c_X$ as a function of the eigenvalue moduli $|\lambda|$. 
We find that, apart from the leading resonance $\lambda_1$, which we highlight with a red star, and some intermediate resonances, plotted as yellow circles, there is a dense cluster of resonances with a nonzero overlap, ranging from $\lambda_{17}$ up to $\lambda_{25}$, whose moduli approximately coincide at $|\lambda_i| \approx 0.40$. These are highlighted with green circles.
In panel $(b)$ of Fig. \ref{Overlaps and spectrum} we show the spectrum plotted on the complex plane, marking with a red star the leading resonance $\lambda_1$, in yellow the intermediate resonances $\lambda_5$ and $\lambda_7$ - $\lambda_{10}$, and in green the resonances $\lambda_{17}$ up to $\lambda_{25}$. 

\begin{figure}[htpb]
    \centering
    \includegraphics[width=1\linewidth]{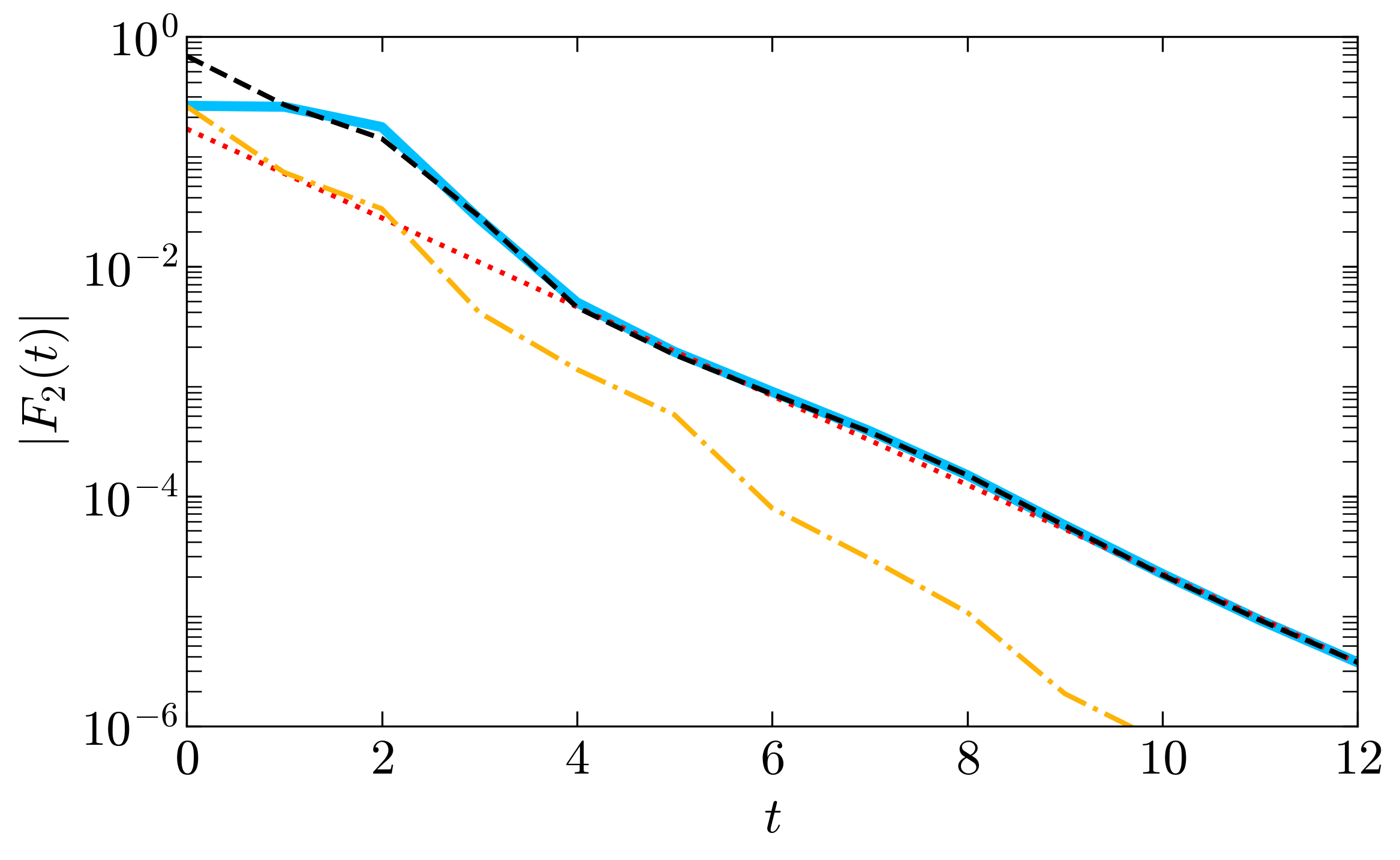}
    \caption{Decay of the coarse-grained 2-OTOC at $\epsilon \approx 0.1$ (blue) together with the exact expansion truncated at $\lambda_1$ (dotted red line) and $\lambda_{25}$ (dashed black line), for the operator $X(t)P$ for the standard map at $K_S = 8$ and Hilbert space dimension $d = 5000$. The yellow, dash-dotted line represents the contribution of the intermediate resonances $\lambda_5$, $\lambda_7$ -- $\lambda_{10}$.}
    \label{fit of coarse-grained 2-OTOC}
\end{figure}

Then we compared the coarse-grained evolution at $\epsilon = 0.1$ and the exact expansion \eqref{exact expansion OTOC} truncated at $\lambda_{25}$. This is shown in Fig.~\ref{fit of coarse-grained 2-OTOC}, where we also plot the contribution of the leading resonance $\lambda_1$. 
We can see that the truncated expansion approximates almost perfectly the 2-OTOC already at times $t \geq 2$, with the leading RPR dominating already at $t \geq 4$.
Moreover, we plot the contribution of the intermediate resonances $\lambda_5$ and $\lambda_7$ -- $\lambda_{10}$. It can be seen that their contribution becomes smaller than that of $\lambda_1$ already at short times $t \geq 2$, and therefore has a negligible contribution.

\bibliography{refs}

\end{document}